\documentclass[aps,twocolumn,a4paper,floatfix,superscriptaddress,pra]{revtex4-1}
\usepackage[toc,page]{appendix}
\usepackage{braket}
\usepackage{epsfig,graphicx,times}
\usepackage{amstext}
\usepackage{amsmath}
\usepackage{amssymb}
\usepackage{mathrsfs}
\usepackage{dcolumn}
\usepackage{bm}
\usepackage[tight]{subfigure}
\usepackage[colorlinks,linkcolor=blue,anchorcolor=blue,citecolor=blue,urlcolor=blue]{hyperref}
\usepackage{color}
\usepackage{siunitx}
\usepackage{appendix}
\usepackage{placeins}
\usepackage{textcomp}
\usepackage{bbold}
\usepackage{float}
\usepackage{verbatim}
\usepackage{minitoc}
\usepackage[dvipsnames]{xcolor}

\usepackage{soul}
\usepackage[framemethod=tikz]{mdframed}
\usepackage{colortbl}  
\usepackage{xcolor}
\usepackage{array}
\begin{document}

\title{Utilizing encoding time as a resource to enhance quantum sensing by probe qubit dephasing}

\author{Ji-Bing Yuan}\email{jbyuan@hynu.edu.cn}
\affiliation{Key Laboratory of Opto-electronic Control and Detection Technology of University of Hunan Province, and College of Physics and Electronic Engineering, Hengyang Normal University, Hengyang 421002, China}

\author{Hai-Fei Liu}
\affiliation{Key Laboratory of Opto-electronic Control and Detection Technology of University of Hunan Province, and College of Physics and Electronic Engineering, Hengyang Normal University, Hengyang 421002, China}

\author{Ya-Ju Song}
\affiliation{Key Laboratory of Opto-electronic Control and Detection Technology of University of Hunan Province, and College of Physics and Electronic Engineering, Hengyang Normal University, Hengyang 421002, China}

\author{Shi-Qing Tang}
\affiliation{Key Laboratory of Opto-electronic Control and Detection Technology of University of Hunan Province, and College of Physics and Electronic Engineering, Hengyang Normal University, Hengyang 421002, China}

\author{Xin-Wen Wang}\email{xwwang@hynu.edu.cn}
\affiliation{Key Laboratory of Opto-electronic Control and Detection Technology of University of Hunan Province, and College of Physics and Electronic Engineering, Hengyang Normal University, Hengyang 421002, China}

\author{Le-Man Kuang}\email{lmkuang@hunnu.edu.cn}
\affiliation{Key Laboratory of Low-Dimensional Quantum Structures and Quantum Control of Ministry of Education, and Department of Physics, Hunan Normal University, Changsha 410081, China}
\affiliation{Synergetic Innovation Academy for Quantum Science and Technology, Zhengzhou University of Light Industry, Zhengzhou 450002, China}
\date{\today}

\begin{abstract}
We examine a system in which an impurity qubit is immersed in a quasi-two-dimensional dipolar Bose-Einstein condensate whose collective excitations act as a depasing reservoir for the qubit. The relative dipole-dipole interaction  strength is estimated by the probe qubit dephasing. The ultimate precision of this estimation is quantified by the quantum Fisher information, which can be obtained by means of measuring quantum coherence of the probe qubit. Our findings indicate that, in the interval where roton excitations appear, the quantum Fisher information oscillates periodically with the encoding time $t$, and the amplitude of these oscillations increases alongside the extension of $t$. Moreover, we analytically determine that the envelope curve formed by the local maximum points satisfies the functional relationship $At+Bt^{1/2}+C$ during long-term encoding scenarios, where $A$, $B$, $C$ are positive numbers. It is also revealed that the highly non-Markovian effects caused by the roton softening of the excitation spectrum allow long encoding time to serve as a resource for enhancing sensing precision. Our work provides a new pathway for enhancing the sensing precision of dephasing qubits.
\end{abstract}


\maketitle
\section{\label{Sec:1}Introduction}
Quantum sensing aims to leverage quantum resources such as coherence~\cite{Adam2022,Mitchison2020}, entanglement~\cite{Nagata2007,Treutlein2018,Haine2020,Planella2022},  and squeezing~\cite{Muessel2014,Bai2019} to improve the precision of various physical measurements. It is essential to accurately sense various characteristic parameters of the environment surrounding a quantum system in quantum control and quantum information processing, as any quantum system inevitably interacts with its surrounding environment, leading to decoherence~\cite{Breuer2007}. However, achieving this is challenging due to the numerous degrees of freedom in the environment. An effective strategy to address this challenge is the utilization of quantum probes, which are small, controllable, and measurable quantum systems~\cite{Benedetti2014,Grasselli2018,Mehboudi2019,Khan2022,Tan2020,Tan2022,Yuan2024,Zhang2024}. A quantum probe is initially prepared in an appropriate state and then placed in the environment to interact. During this interaction, the information about the environment parameters to be estimated encodes into the quantum state of the probe. Finally, measurements are performed on the probe to extract this information. The precision of this estimation has been extensively studied using tools from quantum parameter estimation theory ~\cite{Helstrom1976,Holevo1982}. According to the theory, the ultimate precision of any estimation procedure is constrained by the quantum Cram\'{e}r-Rao (QCR) bound, which can be quantified by the quantum Fisher information (QFI) ~\cite{Wei2013,Jing2019}. A higher QFI indicates greater potential achievable precision.

As a simple and widely used quantum system, the qubits have attracted considerable attention as a quantum probe for sensing environments ~\cite{Jevtic2015,Porras2017,Tamascelli2020,Chu2020,Wolski2020,Shi2020,Blondeau2021,Ai2021,Zheng2022,Aiachei2024}. In many instances, when considering the interaction between the qubits and their environment without energy exchange, the dynamics of the probe can be effectively described by a dephasing model. The dephasing qubits have been widely applied to detect various properties of reservoirs such as measuring ultra-low temperatures~\cite{Razavian2019,Francesca2020,Candeloro2021,Yuan2023} and probing the cutoff frequency and coupling strength of Ohmic reservoirs~\cite{Benedetti2018,Sehdaran2019,Bahrampour2019,Ather2021}. In these sensing processes, the information about the parameter being measured is encoded into the decoherence factor of the qubits' evolving state. It is easy to infer that an initial short period of encoding time can enhance the precision of sensing. And if the coherence of the qubits degrades over extended encoding time, the sensing precision may decrease. In fact, in studies reported on the use of qubits dephasing for quantum sensing, there generally exist an optimal encoding time that maximizes the QFI. Therefore, these sensing schemes require precise control over the measurement time. In this work, we will demonstrate a novel dynamics of QFI when using a single qubit dephasing to sense the environment, suggesting that long encoding time can be used as a resource for enhancing sensing precision.

Owing to their unprecedented controllability and low temperature, atomic Bose-Einstein condensates (BECs) are often referred as the reservoirs suitable for engineering in both experimental~\cite{Palzer2009,Will2011,Spethmann2012,Scelle2013,Zipkes2010,Schmid2010,Balewski2013} and theoretical\cite{Recati2005,Cirone2009,Haikka2011,Haikka2013,Song2019,Yuan2017}studies.
In this work, we consider a system in which an impurity qubit is immersed in a quasi-two-dimensional (quasi-2D) dipolar BEC whose collective excitations act as a depasing reservoir for the qubit. We have shown that, by increasing the relative strength of the dipole-dipole interaction (DDI), the dephasing dynamics of the qubit changed from being Markovian to weak non-Markovian and eventually to highly non-Markovian. Especially, when roton excitations are present, the non-Markovianity becomes divergent~\cite{Yuan2017}.  Meanwhile, recent studies have demonstrated that non-Markovian effects have a positive impact on quantum sensing~\cite{zhou2021,Wu2021,Zhang2022,Xu2023,Aiache2024}. These findings inspire us to use the qubit dephasing for sensing the relative DDI. Our results show that, during the interval where roton excitations are present, the QFI oscillates periodically with encoding time $t$, and the amplitude of these oscillations increases as $t$ extends.  This indicates that extending the encoding time can enhance the precision of quantum sensing. Additionally, we derive an analytical expression for the QFI as a function of time and determine that the envelope curve formed by the local maximum points follows the functional relationship $At+Bt^{1/2}+C$ during long-term encoding scenarios, where $A$, $B$, $C$ are positive numbers.  We further point out that the highly non-Markovian effects caused by the roton softening of the excitation spectrum allow long encoding time to serve as a resource for enhancing sensing precision. This work offers a new approach for improving the sensing precision of dephasing qubits.

The paper is structured as follows: In Sec. II, we present the physical model of a single atomic qubit immersed in a thermally equilibrated quasi-2D dipolar atomic gas reservoir and  propose the sensing scheme to estimate the relative DDI. Section III presents numerical and analytical results for the evolution of the QFI over time, demonstrating that encoding time can serve as a resource to enhance the precision of quantum sensing. Finally, a conclusion is given in Sec. IV.
\section{\label{Sec:2} quantum sensing protocol}
\begin{figure}[tbp]
\includegraphics[width=0.49\textwidth]{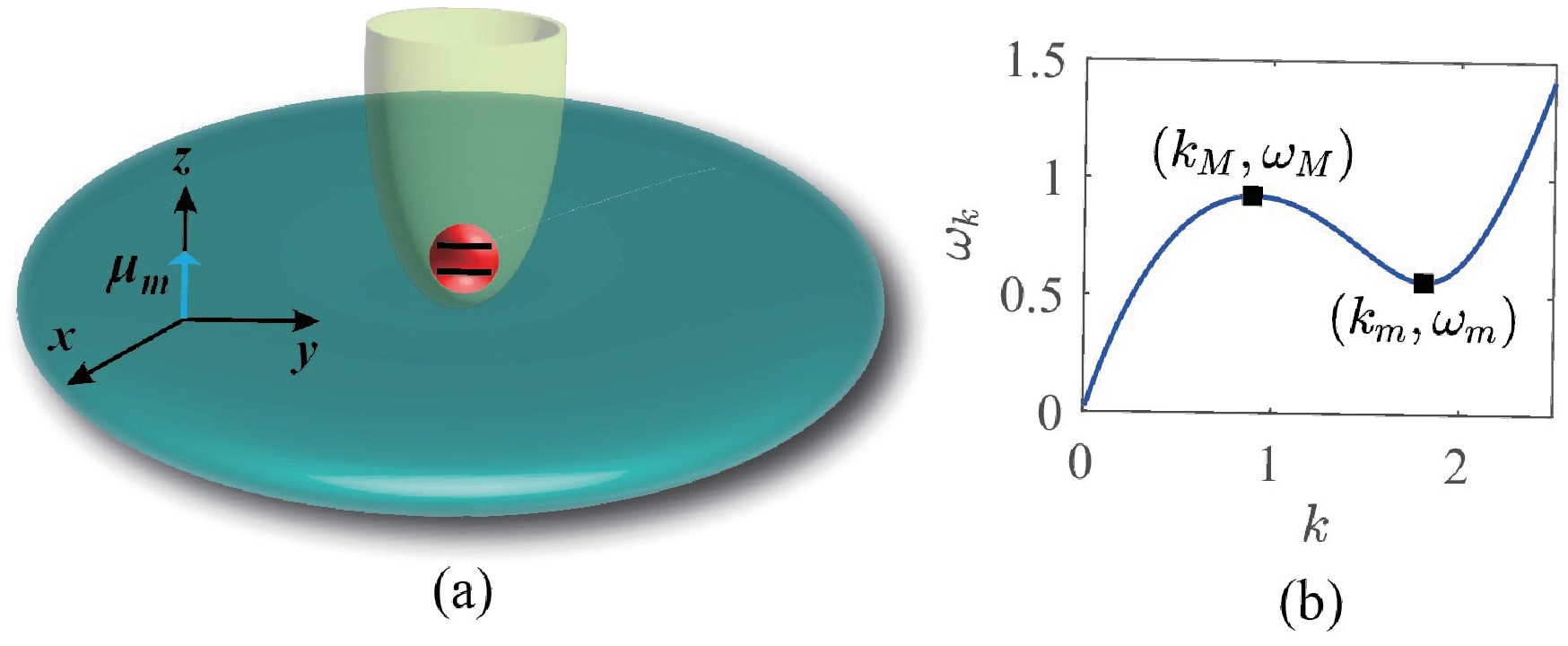}
\caption{(color online) Schematic diagrams of (a) an atomic qubit immersed in
a quasi-2D dipolar BEC and (b) the typical roton spectrum of dipolar
BECs with one minima ($k_{m}$, $\omega_{m}$) and one maxima ($k_{M}$, $\omega_{M}$).} \label{fig1}
\end{figure}
As shown in Fig.~\ref{fig1}(a), we consider a system in which a single atomic qubit is immersed in a thermally equilibrated quasi-2D dipolar atomic gas reservoir at temperature $T$.
The qubit is confined in a harmonic trap $V_A({\bf x})=m_A\omega_{A}^{2}{\bf x}^{2}/2$ that is independent of the internal states, where $m_A$ is the mass of the impurity and $\omega_{A}$ is the trap frequency. For $\hbar\omega_{A}\gg k_{B}T$, the spatial wave function of the qubit is the ground state of $V_{A}(\mathbf{x})$, i.e., $\varphi_{A}({\bf x})=\pi^{-3/4}\ell_{A}^{-3/2}\exp[-{\bf x}^{2}/(2\ell_{A}^{2})]$ with $\ell_{A}=\sqrt{\hbar/(m_{A}\omega_{A})}$. The Hamiltonian of the qubit is $$\hat{H}_{A}=\hbar\Omega_{A}|e\rangle\langle e|,$$ where $\hbar\Omega_{A}$ is level splitting between the ground ($|g\rangle$) and excited ($|e\rangle$) states.

For the reservoir, we assume that each atom has a magnetic dipole moment $\mu_{m}$ polarized in the the $z$ direction. Consequently,  two atoms interact via the potential $$V^{(3D)}\left(\mathbf{x}-\mathbf{x}'\right)=g_{B}\delta(\mathbf{x}-\mathbf{x}')+\frac{3g_{D}}{4\pi}\frac{1-3\cos^{2}\theta}{|\mathbf{x}-\mathbf{x}'|^{3}},$$ where $g_{B}=4\pi \hbar^{2} a_{B}/m_{B}$ represents the contact interaction strength with $m_{B}$ being the mass of the reservoir atom and $a_{B}$ the $s$-wave scattering length, $g_{D}=\mu_{0}\mu_{m}^{2}/3$ with $\mu_{0}$ being the permeability of vacuum, and $\theta$ is the polar angle of ${\mathbf x}-{\mathbf x}'$. Additionally, the gas is confined along the $z$ axis by the potential $V_{B}(z)=m_{B}\omega_{z}^{2}z^{2}/2$, where $\omega_{z}$ is the trap frequency. For sufficiently large $\omega_{z}$, the motion of the atoms along the $z$ axis is frozen to the ground state of $V_{B}(z)$, i.e., $\varphi_{B}(z)=\pi^{-1/4}\ell_{B}^{-1/2}\exp[-z^{2}/(2\ell_{B}^{2})]$ with $\ell_{B}=\sqrt{\hbar/(m_{B}\omega_{z})}$,  effectively reducing the reservoir into a quasi-2D one. Finally, for small $T$, we assume that most of the reservoir atoms are condensed to zero momentum state with an area density $n$.  Following Bogoliubov's method, the uncondensed atoms are then described by the quasiparticle Hamiltonian~\cite{Oosten2001}
\begin{equation}
\hat{H}_B =\sum_{{\bf k}\neq 0}\hbar \omega_{k} \hat{b}^{\dag}_{{\bf k}}\hat{b}_{{\bf k}}\nonumber,
\end{equation}
where ${\mathbf k}\equiv (k_{x},k_{y})$, $\hat b_{\mathbf k}$ is annihilation operator for the quasiparticle with wave vector ${\mathbf k}$, and excitation frequency is~\cite{Fischer2006}
\begin{align}
\omega_{k}=\frac{1}{2}\sqrt{k^{4}+P k^{2}\left[1+\chi\tilde v_{D}(k)\right]}\label{ek}
\end{align}
with $P=8\sqrt{2\pi} \ell_{B}a_{B}n$ being a dimensionless parameter, $\tilde v_{D}(x)=2-3\sqrt{\pi/2}xe^{x^{2}/2}\mathrm{erfc}\left(x/\sqrt{2}\right)$ being the Fourier transform of the effective 2D DDI, and $$\chi=\frac{g_{D}}{g_{B}}$$ being the relative DDI strength. In this paper, we use $\omega_{z}$ as the unit of frequency and $\ell_{B}^{-1}$ as the unit of wave vector. It is now well-established that the sufficiently strong DDI would lead to the roton excitation and eventually the instability. In fact, for $P=2$, roton excitation sets in when $$\chi>\chi^{*}\simeq 4.23.$$ In addition, the condensate becomes unstable for $\chi>\chi^{**}=5.67$. The typical roton spectrum is shown in Fig.~\ref{fig1}(b).

For the qubit-reservoir coupling, we assume that the qubit undergoes $s$-wave collisions with reservoir atoms only when the qubit is in the excited state~\cite{Recati2005,Cirone2009,Haikka2011,Haikka2013}. Let $a_{AB}$ be the corresponding scattering length, the qubit-reservoir interaction Hamiltonian is then
\begin{equation}
\hat{H}_{AB}=\hbar\delta_{e}|e\rangle\langle e|+\hbar|e\rangle\langle e|
\sum_{\mathbf{k}\neq 0}g_{k}\left(\hat{b}_{\mathbf{k}}+\hat{b}^{\dag}_{\mathbf{k}}\right)\nonumber,
\end{equation}
where $\delta_{e}=2\sqrt{\pi}na_{AB}m_{B}\ell^{2}_{B}/[m_{AB}(\ell^{2}_{A}+\ell^{2}_{B})^{1/2}]$ is the excited level shift due to the collision, $m_{AB}=m_{A}m_{B}/(m_{A}+m_{B})$ is the reduced mass, and
 \begin{equation}
g_{k}=\frac{\delta_e}{\sqrt{2nS}}\frac{k\mathrm{e}^{-\frac{\zeta^{2}k^{2}}{4}}}{\sqrt{\omega_{k}}}\label{gk}
\end{equation}
is the qubit-reservoir coupling parameter, with $S$ being the area of the reservoir and $\zeta=\ell_{A}/\ell_{B}$.

Now the total Hamiltonian, $\hat H=\hat H_{A}+\hat H_{B}+\hat H_{AB}$, is
\begin{equation}
\hat{H}=\hbar(\Omega_{A}+\delta_{e})|e\rangle\langle
e|+\sum_{{\mathbf k}\neq 0}\hbar\omega_{k}\hat{b}^{\dag}_{\mathbf k}\hat{b}_{\mathbf k}+\sum_{\mathbf{k}\neq0}\hbar g_{k}|e\rangle\langle e|\left(\hat{b}_{\mathbf{k}}+\hat{b}^{\dag}_{\mathbf{k}}\right),\label{hami}
\end{equation}
Since $\hat H_{A}$ commutes with $\hat H_{AB}$, the dynamics of the impurity qubit in reservoir is purely dephasing. In the following we use the probe qubit dephasing to estimate the relative DDI strength $\chi$ and put forward the following quantum sensing protocol: (i) First we initialize the probe qubit in the superposition state $|\psi\rangle=(|g\rangle +|e\rangle)/\sqrt{2}$. (ii) Then the probe qubit undergoes a parameter-encoding process by interacting with the dipolar BEC. (iii) Finally, we perform a measurement $\hat{\sigma}_{x}$ on the probe qubit.

We now detail the sensing protocol. The initial state of the total system is prepared as
\begin{equation}
\hat{\rho}_{tot}(0)=|\psi\rangle\langle\psi|\otimes\hat{\rho}_{B} \nonumber,
\end{equation}
where $\hat{\rho}_{B}=\prod_{\mathbf k}\left(1-\mathrm{e}^{\beta\omega_{k}}\right)\mathrm{e}^{-\beta\omega_{k}b_{{\mathbf k}}^{\dag }b_{{\bf k}}}$ is a thermal state of the BEC with $\beta=1/k_{B}T$. For simplicity, we shall only consider the zero temperature case in this work. Let the whole system evolve under the control of the Hamiltonian (\ref{hami}) for a certain time $t/2$, after which a $\pi$-pulse about $x$ is applied to the qubit. Then the system is allowed to evolve for the same time period $t/2$ and another $\pi$-pulse is applied. Through these processes, the quantum state of the probe qubit at time $t$ can be given as
\begin{equation}
\hat{\rho}_{A}(t)=\frac{1}{2}\left(
                \begin{array}{cc}
                  1 &\mathrm{e}^{-\Gamma(t)} \\
                  \mathrm{e}^{-\Gamma(t)} & 1 \\
                \end{array}
              \right)
,\label{state}
\end{equation}
where $\Gamma(t)$ is the decoherence factor with following expression
\begin{equation}
\Gamma(t)=\sum_{\mathbf{k}\neq0}\frac{g_{k}^{2}}{\omega_{k}^{2}}\left[1-\cos(\omega_{k}t)\right].\label{gama1}
\end{equation}
Here and henceforth, we adopt $\omega_{z}^{-1}$ as the unit of time.

From Eq.~(\ref{ek}) and Eq.~(\ref{gama1}), we see that information of the relative DDI strength $\chi$ is encoded into the decoherence factor of the qubit. In the following, we introduce the quantum parameter estimate theory to quantify the sensing precision. As is well-known, the sensing precision of $\chi$ is restricted to the QCR bound
 \begin{equation}
\delta \chi \geq \frac{1}{\sqrt{\nu  \mathcal{F}^{Q}_{\chi}}}.\label{QCR}
\end{equation}
 Here $\delta \chi$ is the mean square error, $\nu$ represents the number of repeated experiments and $\mathcal{F}^{Q}_{\chi}$ denotes QFI with respect to the $\chi$. The QCR bound in Eq.~(\ref{QCR}) indicates that a larger amount of QFI corresponds to enhanced potential sensing precision. The QFI for state ~(\ref{state}) is given as~\cite{Wei2013,Jing2019}
\begin{equation}
\mathcal{F}^{Q}_{\chi}=\frac{\left(\partial_{\chi}\Gamma\right)^{2}}{\mathrm{e}^{2\Gamma}-1}.\label{fis}
\end{equation}
In this paper, $\partial_{x}Y$ represents the partial derivative of $Y$ with respect to $x$.

The final stage of the sensing protocol involves determining the optimal measurement $\hat{\Lambda}$ which can saturate the QCR bound. For a qubit system, the Fisher information associated with the measurement $\hat{X}$ can be presented as~\cite{Mitchison2020}
\begin{equation}
\mathcal{F}_{\chi}=\frac{\left(\partial_{\chi}\langle\hat{X}\rangle\right)^{2}}{\langle\Delta \hat{X}^{2}\rangle}\nonumber,
\end{equation}
where $\langle\hat{X}\rangle$ and $\langle\Delta \hat{X}^{2}\rangle$ are mean and variance of the measured observable, respectively. The QFI is the upper bound of the Fisher information associated with the measurement, i.e.,
\begin{equation}
 \mathcal{F}^{Q}_{\chi}=\max_{\hat{X}}\mathcal{F}_{\chi}(\hat{X})=\mathcal{F}_{\chi}(\hat{\Lambda})\nonumber.
\end{equation}
It can be proven that the Fisher information associated with the pauli matrices $\hat{\sigma}_{x}$ is exactly equal to the QFI given by Eq.~(\ref{fis})~\cite{Razavian2019,Yuan2023}. Therefore, we shall choose $\langle\hat{\sigma}_{x}\rangle$ as the measurement signal, which can be observed using Ramsey interferometry~\cite{Adam2022,Scelle2013,Cetina2016}.

\section{\label{Sec:3} Quantum sensing enhancement via long encoding time}
\subsection{ Numerical results}
In this subsection, we begin by examining the evolution of the decoherence factor over time. To achieve this, we perform a continuization process on the decoherence factor in Eq.~(\ref{gama1}). Substituting $g_{k}$ in Eq.~(\ref{gk}) into Eq.~(\ref{gama1}) and using the continuum limit $\frac{1}{S}\sum_{\mathbf{k}}\rightarrow\frac{1}{2\pi\ell^{2}_{B}}\int^{\infty}_{0}k \mathrm{d}k$, the decoherence factor can be rewritten as
\begin{eqnarray}\label{gama2}
\Gamma(t)=Q \int^{\infty}_{0}f(k)\frac{\left[1-\cos(\omega_{k}t)\right]}{\omega_{k}^{3}}\mathrm{d}k
\end{eqnarray}
where $Q=na^{2}_{AB}\ell_{B}^{2}\left(m_{A}+m_{B}\right)^{2}/\left[m_{A}^{2}\left(\ell_{A}^{2}+\ell_{B}^{2}\right)\right]$ is a dimensionless parameter measuring the qubit-reservoir coupling and $f(k)\equiv k^{3}\mathrm{e}^{-\zeta^{2}k^{2}/2}$. Combining Eqs.~(\ref{ek}) and~(\ref{gama2}), we know that the decoherence factor $\Gamma$ is a function of $t$, $\chi$, $P$, and $Q$. To fix $P$ and $Q$, we consider a single $^{87}$Rb atom immersed in a dipolar BEC $^{164}$Dy atoms~\cite{Lu2011} which possess a magnetic dipole moment of $10\mu_{B}$. We assume a typical trap frequency $\omega_{z}=2\pi\times 10^{3}\,{\rm Hz}$, the corresponding harmonic oscillator width is $\ell_{B}\simeq 2.5\times 10^{-5}\,{\rm cm}$. Next, we consider a typical condensate peak density of $10^{14}\,{\rm cm}^{-3}$, the area density is then $n=4.4\times 10^{9}\,{\rm cm}^{-2}$. Consequently, we have $P\sim 1.5$. To find $Q$, we assume that the $s$-wave scattering length between $^{87}$Rb and $^{164}$Dy atoms is $a_{AB}\sim 5\,{\rm nm}$ and the width of the impurity trap $\ell_{A}$ equals to $\ell_{B}$ ($\zeta=1$), we therefore have $Q\sim 4.6\times 10^{-3}$. Without loss of generality, we shall take $P=2$ and $Q=4\times10^{-3}$ in the results presented below. It has been verified that changing the values of $P$ and $Q$ would not change our main results qualitatively.

\begin{figure*}[tbp]
\centering
\includegraphics[clip=true,height=6cm,width=16cm]{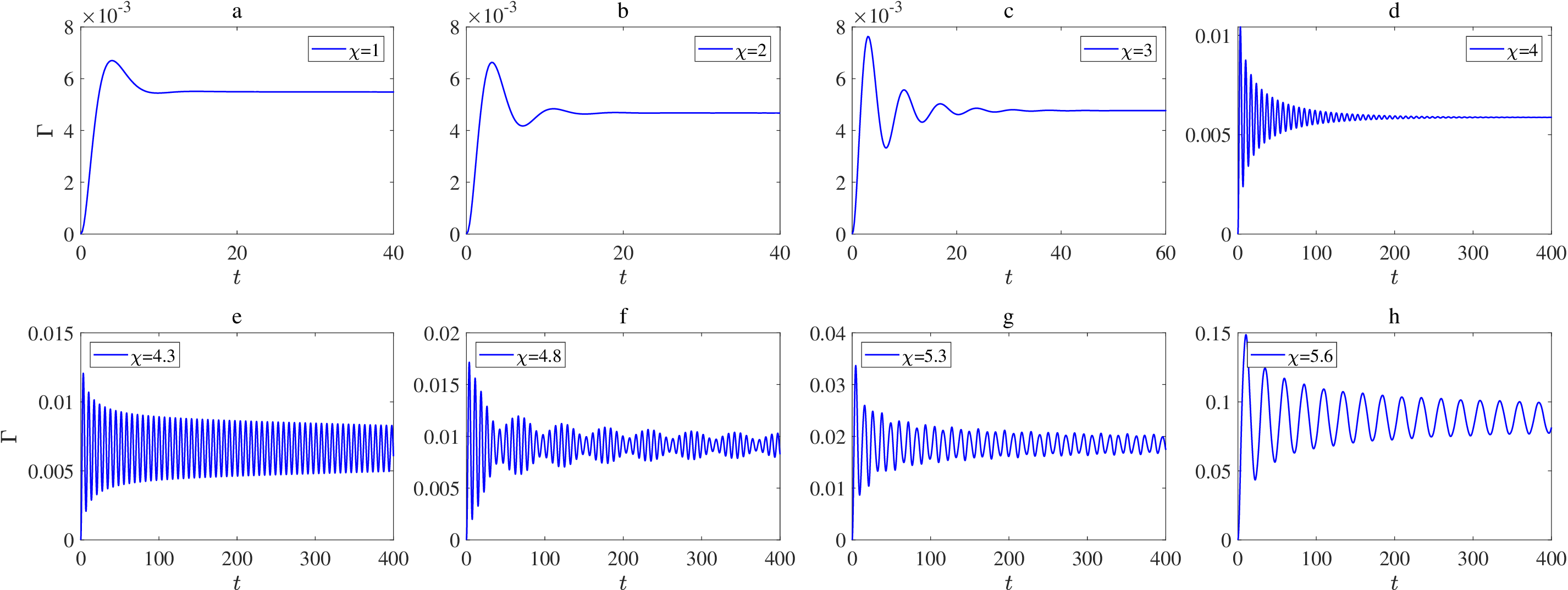}
\caption{(color online)Time dependence of the decoherence factor $\Gamma$ for $\chi=1$ (a), $2$ (b), $3$ (c),$4$ (d), $4.3$ (e), $4.8$ (f), $5.3$ (g), and $5.6$ (h).} \label{garm}
\end{figure*}

Based on Eq.~(\ref{gama2}), we plot the typical dynamical behaviors of $\Gamma(t)$ for various values of $\chi$ in Fig.~\ref{garm}. From the top four subfigures, we can observer that as $\chi$ increases, the number of oscillations of $\Gamma$ over time increases; but they all quickly converge to the asymptotic value after a few oscillations. The bottom four subfigures show us that when $\chi>\chi^{*}$ where the roton excitation appear, $\Gamma(t)$ becomes a damped oscillating function that oscillates for a very long period of time. The fact that $\Gamma(t)$ exhibits very distinct behaviors for different $\chi$ inspires us to use the qubit dephasing to estimate the relative DDI $\chi$.

\begin{figure*}[tbp]
 \centering
 \includegraphics[clip=true,height=6cm,width=16cm]{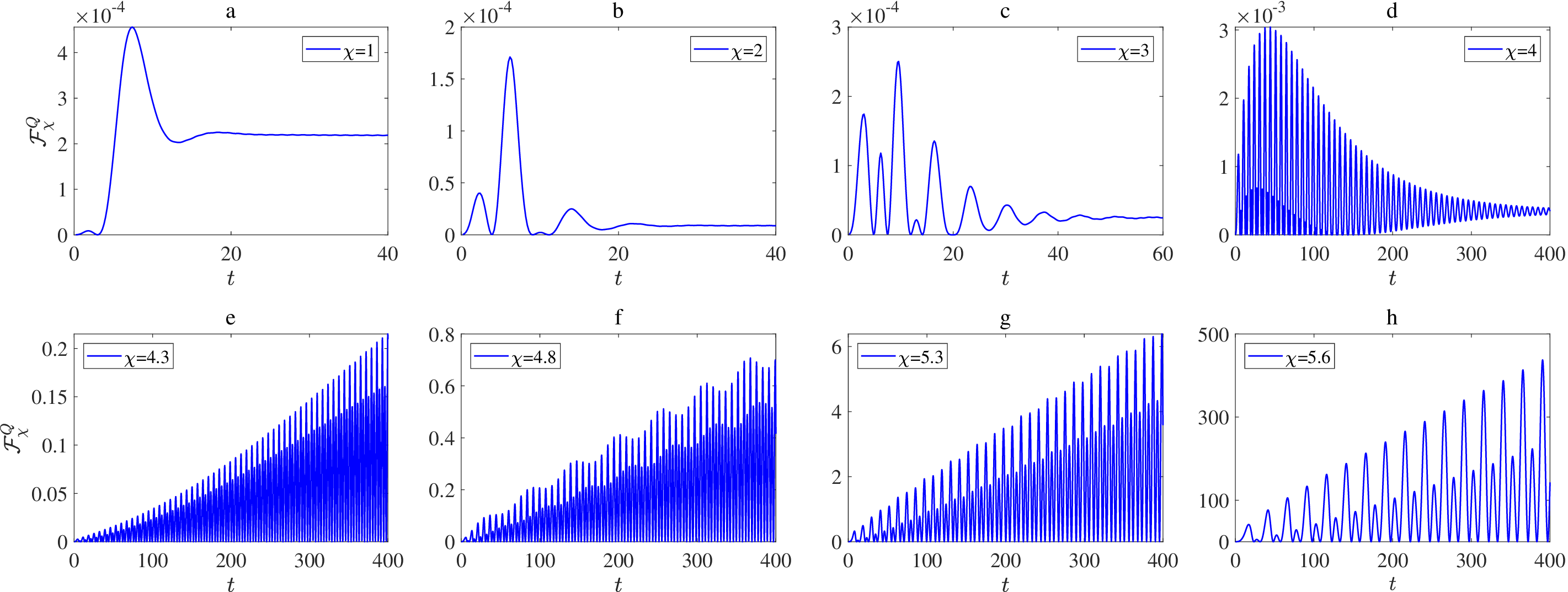}
\caption{(color online) Time dependence of the QFI $\mathcal{F}^{Q}_{\chi}$ for $\chi=1$ (a), $2$ (b), $3$ (c),$4$ (d), $4.3$ (e), $4.8$ (f), $5.3$ (g), and $5.6$ (h).} \label{fisher}
\end{figure*}

 Figure~\ref{fisher} shows the dynamical behaviors of $\mathcal{F}^{Q}_{\chi}$ corresponding to the values of $\chi$ in Fig.~\ref{garm}.  The top four subfigures  demonstrate that there exist an optimal encoding time at which the QFI reaches its maximum, similar to the QFI dynamics studied previously~\cite{Razavian2019,Francesca2020,Candeloro2021,Yuan2023,Benedetti2018,Sehdaran2019,Bahrampour2019,Ather2021}. However, when $\chi>\chi^{*}$ ,  the bottom four subfigures show that the QFI oscillates periodically with the encoding time $t$, and the amplitude of these oscillations increases alongside the extension of $t$. In fact, we numerically find that this growth trend still holds for $t$ over $1000$. Moreover, by comparing the bottom four subfigures, we observe that the local maximum attainable value of QFI within the same time period increases as $\chi$ increases. Notably, when $\chi$ approaches $\chi^{**}$, this maximum value rises sharply. These dynamical behaviors suggest that extending the encoding time can enhance  quantum sensing  precision  within the range where roton excitations appear. And the degree of enhancement becomes more pronounced as $\chi$ is close to the unstable point $\chi^{**}$.

\subsection{ Analytical findings}
In this subsection, to gain a better understanding of why QFI increases with encoding time in the interval $\chi>\chi^{*}$, we attempt to derive an approximate analytical expression for the evolution of QFI over time.
To achieve this, we divide $\Gamma(t)$ in Eq.~(\ref{gama2}) into two parts: the time-independent part $$\Gamma_{0}=Q \int^{\infty}_{0}\frac{f(k)}{\omega_{k}^{3}}\mathrm{d}k$$ and the time-dependent part
\begin{equation}
\Gamma_{1}(t)=-\int_{0}^{\infty}G(\omega)\cos(\omega t)\mathrm{d}\omega,\label{gamat}
\end{equation}
where $G(\omega)$ is given as
\begin{align}
G(\omega)=Q\sum_{i}\frac{f(k_{i}(\omega))}{\omega^{3}}\left|\frac{\mathrm{d}\omega_{k}}{\mathrm{d}k}\right|^{-1}_{k={k_{i}(\omega)}}\label{spec}
\end{align}
with $k_{i}(\omega)$ being the root of the equation $\omega_{k}=\omega$. As shown in Fig.~\ref{fig1}(b), the roton spectrum has a minima ($k_{m}$, $\omega_{m}$) and a maxima ($k_{M}$, $\omega_{M}$). Based on Eq.~(\ref{spec}), $G(\omega)$ diverges at $\omega_{m}$ and $\omega_{M}$. To accurately take into account the contributions from these singularities to $G(\omega)$, let us focus on $\omega_{k}$ in the vicinities of $k_{m}$ where the excitation energy can be approximated as $\omega_{k}\approx \hbar\omega_{m}+\omega_{k}''(k_{m})(k-k_{m})^{2}/2$. Using Eq.~(\ref{spec}), it can be then shown that, in the vicinity of $\omega_{m}$, we have $G(\omega)\approx g_{m}(\omega_{m}-\omega)^{-1/2}$ for $\omega>\omega_{m}$, where $$g_{m}=Q\sqrt{\frac{2}{\omega_{k}^{''}(k_{m})}}\frac{f(k_{m})}{\omega_{m}^{3}}.$$ Similarly, in the vicinity of the maxima, we have $G(\omega)\approx g_{M}(\omega-\omega_{M})^{-1/2}$ for $\omega<\omega_{M}$, where$$g_{M}=Q\sqrt{\frac{2}{\left|\omega_{k}^{''}(k_{M})\right|}}\frac{f(k_{M})}{\omega_{M}^{3}}.$$ Now, by assuming that these two singularities give rise to the largest contribution to $G(\omega)$, we define function
\begin{align}
\tilde G(\omega)=g_{m}\frac{H(\omega-\omega_{m})}{\sqrt{\omega-\omega_{m}}}+g_{M}\frac{H(\omega_{M}-\omega)}{\sqrt{\omega_{M}-\omega}}\label{approxg}
\end{align}
as the approximate of $G(\omega)$, where $H(x)$ is the Heaviside step function. By substituting $\tilde G(\omega)$ for $G(\omega)$ in Eq.~(\ref{gamat}), we obtain an approximate $\Gamma_{1}(t)$
\begin{equation}\label{apgamat}
\tilde{\Gamma}_{1}(t)=-\sqrt{\frac{\pi}{t}}\left[g_{m}\cos\left(\omega_{m} t+\frac{\pi}{4}\right)+g_{M}\cos\left(\omega_{M} t-\frac{\pi}{4}\right)\right].
\end{equation}
\begin{figure}[htbp]
\includegraphics[width=0.4\textwidth]{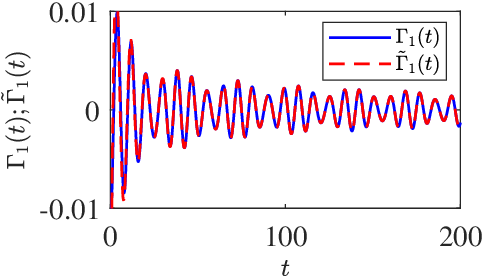}
\centering
\caption{(Color online). Comparison of $\Gamma_{1}(t)$ and $\tilde\Gamma_{1}(t)$ for $\chi=4.8$. Parameters used are $\omega_{m}=0.7971$, $\omega_{M}=0.9114$, $g_{m}=9.5\times10^{-3}$ and $g_{M}=3.8\times10^{-3}$, where $g_{m}$ and $g_{M}$ are expressed in unit of $\omega_{z}^{-1/2}$. } \label{fitg}
\end{figure}
The near-perfect overlap of curves $\Gamma_{1}(t)$ and  $\tilde{\Gamma}_{1}(t)$ in Fig.~\ref{fitg} indicates that the approximation is valid. In fact, including other values of $\chi>\chi^{*}$ , even if the time is extended to $1000$, we find that such overlap would still be satisfied. The variations of $\omega_{m}, \omega_{M}$ and $g_{m}, g_{M}$ with $\chi$ are presented in Fig.~\ref{gomega} (a) and~\ref{gomega} (b), respectively. The decrease in $\omega_{m}$ with increasing $\chi$ observed in Fig.~\ref{gomega} (a) is termed as the roton mode softening~\cite{Fischer2006}.
\begin{figure}[htbp]
\centering
\includegraphics[clip=true,height=4cm,width=8cm]{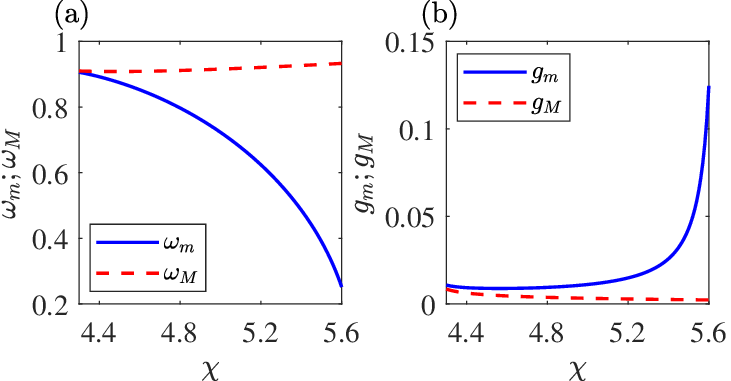}
\caption{(Color online).  $\omega_{m}$ , $\omega_{M}$ in (a) and $g_{m}$, $g_{M}$ (in unit of $\omega_{z}^{-1/2}$) in (b) versus $\chi$.} \label{gomega}
\end{figure}

\begin{figure}[htbp]
\includegraphics[clip=true,height=4cm,width=8cm]{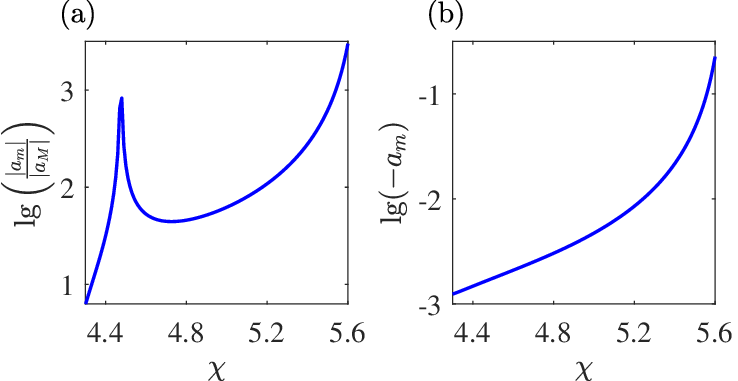}
\caption{(Color online). Plot of $\lg\left(\frac{|a_{m}|}{|a_{M}|}\right)$ in (a) and  $\lg(-a_{m})$ in (b) as a function of $\chi$, where $a_{m}$ and $a_{M}$ are expressed in unit of $\omega_{z}^{1/2}$.  } \label{am}
\end{figure}

Next, based on $\tilde{\Gamma}_{1}(t)$ in Eq.~(\ref{apgamat}), we derive the analytical expression of QFI as a function of time for long-term encoding. Under long-term conditions ($t>>1$), neglecting the decay term related to time, we can obtain $e^{2\Gamma}\approx e^{2\Gamma_{0}}$ and $$\partial_{\chi}\tilde{\Gamma}_{1}\approx\sqrt{\pi t}\left[a_{m}\sin\left(\omega_{m} t+\frac{\pi}{4}\right)+a_{M}\sin\left(\omega_{M} t-\frac{\pi}{4}\right)\right],$$
where $a_{i}=g_{i}\partial_{\chi}\omega_{i}$ with $i=m,M$. From Fig.~\ref{am} (a), it is evident that $|a_{m}|>>|a_{M}|$, allowing for the reasonable approximation as follows
\begin{equation}\label{dgama}
\partial_{\chi}\tilde{\Gamma}_{1}\approx\sqrt{\pi}a_{m}\sqrt{t}\sin\left(\omega_{m} t+\frac{\pi}{4}\right).
\end{equation}
Due to $\partial_{\chi}\omega_{m}<0$ and $g_{m}>0$, $a_{m}$ has to be negative. The change in $\lg(-a_{m})$ as a function of  $\chi$ is illustrated in Fig.~\ref{am} (b).
Using Eq.~(\ref{dgama}), it is straightforward to derive an approximate expression for QFI in Eq.~(\ref{fis})
\begin{equation}\label{apfis}
\tilde{\mathcal{F}}^{Q}_{\chi}=At\sin^{2}\left(\omega_{m} t+\frac{\pi}{4}\right)-B\sqrt{t}\sin\left(\omega_{m} t+\frac{\pi}{4}\right)+C,
\end{equation}
where $A$, $B$, $C$ are time-independent parameters with following forms
\begin{equation}
 A=\frac{\pi a_{m}^{2}}{\mathrm{e}^{2\Gamma_{0}}-1},\hspace{0.3cm}  B=\frac{2\sqrt{\pi}a_{m}\partial_{\chi}\Gamma_{0}}{1-\mathrm{e}^{2\Gamma_{0}}},
  \hspace{0.3cm}  C=\frac{(\partial_{\chi}\Gamma_{0})^{2}}{\mathrm{e}^{2\Gamma_{0}}-1}. \nonumber
\end{equation}
\begin{figure}[htbp]
\includegraphics[clip=true,height=4cm,width=5cm]{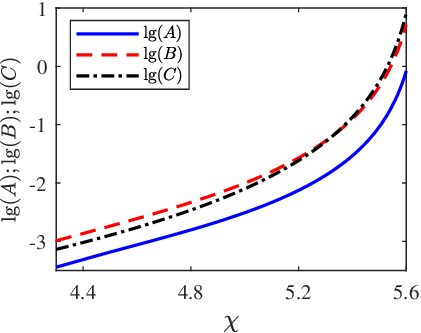}
\caption{(color online). Plot of $\lg(A)$, $\lg(B)$ and $\lg(C)$ as a function of $\chi$. Here the unit for $A$ is $\omega_{z}$, and the unit for $B$ is $\omega_{z}^{1/2}$. }\label{abc}
\end{figure}
The relationship between  $\lg(A)$, $\lg(B)$, $\lg(C)$, and $\chi$ is depicted in Fig.~\ref{abc}. Clearly, $A$, $B$ and $C$ are all positive values. According to Eq.~(\ref{apfis}), it is easy to determine the local optimal time $$t_{LO}=\frac{\left(2n+\frac{5}{4}\right)\pi}{\omega_{m}}$$ at which the QFI reaches a local maximum, where $n$ is a positive integer. The corresponding local maximum of the QFI reads
\begin{eqnarray}\label{apmaxfis}
\tilde{\mathcal{F}}^{Q}_{\chi}(t_{LO})=At_{LO}+B\sqrt{t_{LO}}+C.
\end{eqnarray}
\begin{figure}[htbp]
\includegraphics[clip=true,height=8cm,width=7cm]{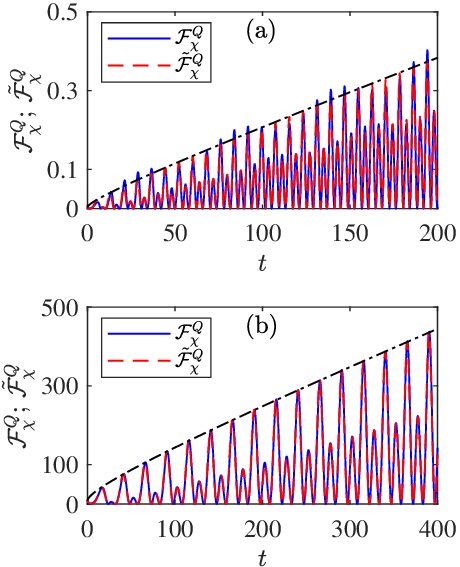}
\caption{(Color online). Comparison of $\mathcal{F}^{Q}_{\chi}$ and  $\tilde{\mathcal{F}}^{Q}_{\chi}$ for $\chi=4.8$ (a) and $\chi=5.6$ (b). Parameters used in (a) are $\omega_{m}=0.7971$, $A=1.6\times10^{-3}$, $B=4.6\times10^{-3}$, and $C=3.4\times10^{-3}$ and in (b) are $\omega_{m}=0.2515$, $A=0.8316$, $B=5.2028$, and $C=8.1374$. }\label{fitf}
\end{figure}
Figure~\ref{abc} illustrates that $A$, $B$, and $C$ all increase by three orders of magnitude from $\chi=4.3$ to $\chi=5.6$, which exactly confirms the observation in Figs.~\ref{fisher} (e) to~\ref{fisher} (h) that the local maximum QFI in the same duration also increases by three orders of magnitude. Finally, to verify the validity of Eq.~(\ref{apfis}), we pick two point $\chi=4.8$ in Fig.~\ref{fitf} (a) and $\chi=5.6$ in Fig.~\ref{fitf} (b) to compare the dynamical behaviors of $\mathcal{F}^{Q}_{\chi}$  and $\tilde{\mathcal{F}}^{Q}_{\chi}$. The plot reveals that these two curves (blue solid line and red dashed line) are nearly identical, achieving an even greater congruence at $\chi=5.6$. The black dash-dotted lines shown in Figs.~\ref{fitf} (a) and~\ref{fitf} (b) are governed by Eq.~(\ref{apmaxfis}), which clearly indicates that the long encoding time can be a resource for enhancing quantum sensing precision.

\subsection{Discussion}
 In this subsection, let's discuss the relationship between non-Markovian effects and the enhancement of quantum sensing via long encoding time.  The non-Markovian effects refer to phenomena in which information flows from the environment back to the system, allowing the open quantum system to recover some of its lost memory~\cite{Breuer2007}. Recent studies have shown that non-Markovian effects positively impact quantum sensing~\cite{zhou2021,Wu2021,Zhang2022,Xu2023,Aiache2024}. For a dephasing qubit, the non-Markovian effects are manifested in the reduction of the decoherence factor over time.  Based on reference~\cite{Breuer2009}, the measure of non-Markovianity is $$\mathscr{N}=\int_{\Gamma'(t)<0} \mathrm{d}\mathrm{e}^{-\Gamma(s)},$$ where the integration is over all intervals in which $\Gamma'(t)<0$. For a Markovian process, the decoherence factor either increases over time or increases to a constant value. These scenarios correspond to the QFI decreasing to zero with prolonged encoding time and remaining constant, respectively. Therefore, we can say that the presence of non-Markovian effects is a necessary condition for extending encoding time to enhance quantum sensing precision. In the interval $\chi<\chi^{*}$,  the presence of non-Markovian effects for only a limited duration, as shown from Figs.~\ref{garm} (a) to~\ref{garm} (d), does not make long encoding time a resource for enhancing quantum sensing precision. This is confirmed by the dynamical behaviors of $\mathcal{F}^{Q}_{\chi}$ in Figs.~\ref{fisher} (a) to~\ref{fisher} (d).  Fortunately, based on Eq.~(\ref{apgamat}), it is found that in the interval $\chi>\chi^{*}$ where roton excitations appear, the non-Markovian effects exhibit the following two special properties: (i) The non-Markovian effects are present throughout the dephasing dynamics, implying that information is exchanged between the qubit probe and the dipolar BEC over a prolonged duration. (ii) The non-Markovianity $\mathscr{N}$ is divergent. This can be proven using the inequality $$\mathscr{N}>\epsilon\sum_{n=1}^{\infty}\frac{1}{\sqrt{n}},$$ where $\epsilon$ is a positive real number and infinite series $\sum_{n=1}^{\infty}1/\sqrt{n}$ is divergent. We believe that such highly non-Markovian effects enable long encoding time to serve as a resource for enhancing sensing precision
 .
\section{\label{Sec:5} Conclusion}
In this work, we considered a system where an impurity qubit is embedded in a quasi-2D dipolar BEC. The collective excitations of the diploar BEC act as a dephasing reservoir for the qubit. By measuring the single-qubit dephasing, we estimated the relative strength of the DDI whose estimation precision is quantified by the QFI. We carefully examined the dynamical behaviors of the QFI at various values of $\chi$. It has been shown that, in the region  $\chi>\chi^{*}$ where roton excitations appear, the QFI oscillates and increases over time. This indicates that extending the encoding time can enhance the precision of quantum sensing. We derived a simple analytical expression for the time evolution of QFI in $\chi>\chi^{*}$, thereby determining that the envelope curve formed by the local maximum points satisfies the functional relationship $At+Bt^{1/2}+C$ during long-term encoding scenarios. It was found that $A$, $B$, and $C$ are all positive numbers and all increased by three orders of magnitude as $\chi$ varied from 4.3 to 5.6. Particularly, as $\chi$ approaches $\chi^{**}$, the rate of increase becomes significantly steeper. It was found that the non-Markovian effects caused by the roton softening of the excitation spectrum have two special properties: (i) The non-Markovian effects are present throughout the dephasing dynamics. (ii) The non-Markovianity $\mathscr{N}$ is divergent. We believe that such highly non-Markovian effects enable long encoding time to serve as a resource for enhancing sensing precision.

Finally, we note that roton-like dispersions are also found in atomic condensates placed inside an optical cavity~\cite{Mottl2012} or with spin-orbit coupling~\cite{Ji2015}. Therefore, when utilizing dephasing qubits to estimate certain parameters of these systems, long encoding time may also serve as a resource for enhancing quantum sensing precision. Our work provides a new pathway for enhancing the sensing precision of dephasing qubits.

\acknowledgments
J. B.  Yuan was supported by  NSFC (No. 11905053) and Scientific Research Fund of Hunan Provincial Education Department of China under Grant (No. 21B0647). L. M. Kuang was supported by NSFC   (Nos. 12247105, 1217050862, and 11935006) and the science and technology innovation Program of Hunan Province (No. 2020RC4047). Y. J. Song was supported by  NSFC (No. 12205088) and Scientific Research Fund of Hunan Provincial Education Department of China under Grant (No. 21B0639). S. Q. Tang was supported by Scientific Research Fund of Hunan Provincial Education Department of China under Grant (No. 22A0507).

\end{document}